# Molecular dynamics analysis of adsorption process of anti-copper-corrosion additives to the copper surface

Kohei Nishikawa[1], Hirotoshi Akiyama[1], Kazuhiro Yagishita[2], Hitoshi Washizu[1, 3*]

[1] Graduate School of Simulation Studies, University of Hyogo, 7-1-28 Minatojima-minamimachi, Chuo-ku, Kobe, Hyogo 650-0047, JAPAN.
[2] JXTG Nippon Oil & Energy Corporation, 8 Chidoricho, Naka-ku, Yokohama, Kanagawa 231-7153, JAPAN.
[3] Elements Strategy Initiative for Catalysts and Batteries (ESICB), Kyoto University - 1-30 Goryo-Ohara, Nishikyo-ku, Kyoto 615-8245, JAPAN.
*Corresponding author: h@washizu.org

of BTA molecules adsorbed on the Cu area is 5 times greater than that on the $Cu_2O$ area. Detailed dynamics focused on charge transfer showed a surface diffusion and enhancement of polarization due to charge transfer from the metal surface caused the selective adsorption. In a real phenomenon, the reason why a few anti-corrosion additives are able to protect a metal surface is postulated to be due to this selective nature of adsorption onto a newly formed metal surface.

## 1.0 INTRODUCTION

Newly formed metal surfaces are often unstable and become stable when they are terminated with other molecules, however, the original color and properties may be diminished when they are covered with oxygen or other gases or liquids in the atmosphere. Anti-copper-corrosion additives adsorb onto the surface of copper and are used in order to prevent these phenomena in order to retain its color and other properties. Benzotriazole (BTA, $C_6H_5N_3$) is one of the famous anti-copper-corrosion additives, and BTA has been used to prevent corrosion and discoloration of copper and copper alloys for a long time (Cotton and Scholes, 1967). There are a huge number of studies concerning this phenomenon (see reviews by Antonijević and Petrović, 2008, Petrović Mihajlović and Antonijević, 2015), and they are applied to novel systems such as ionic liquids (Liu et al., 2006, Zhang et al., 2017), but the mechanism of protecting surfaces is still under study.

From an experimental point of view, the adsorption process of molecules can be analyzed by a surface technique. Recently, a real-time instrumentation technique using Otto-SPR was proposed, and it is becoming possible to observe how additives adsorb onto a surface on a molecular level (Maegawa et al., 2016). Surface-enhanced Raman spectroscopy is also a well-established technique to investigate the surface adsorption process of additives (Ali et al., 2017).

The mechanism is also studied by computational chemistry. In the beginning, a molecular orbital (MO) simulation is used to calculate the partial charges, dipole moment, HOMO-LUMO energy gap, and free volume of the single additive molecule in order to consider the relation with experiments (Edwards et al., 1994). The placement of the additive molecule on the copper surface is studied using a Monte Carlo simulation (Bartley et al., 2003). In this simulation, the dependence of the surface binding energy with an alkyl chain length on alkyl esters of 5-carboxybenzotriazole are obtained. In the Monte Carlo simulation, the classical force field in which the partial charges are fixed is used.

In the adsorption process, electron transfer between solid surface and additive molecule may be important. Density functional theory (DFT) calculations are useful to calculate the interaction between the solid slab and the additive molecule. Using DFT calculation, Jiang et al., 2004 found that the standing, i.e., the plane of the double ring of the BTA molecule is perpendicular to the slab, structure is stable on the $Cu_2O$ slab. The other calculation also shows that the N-Cu bond and intermolecular aggregate structure are important for adsorption of the BTA on the copper surface (Kokalj et al., 2010, Kokalj and Peljhan, 2015). In these studies, since the simulation cell is set to a minimum size to calculate the slab-additive structure, the "aggregate" means that the interactions are calculated between single molecules in the periodic boundary. DFT calculations are used in other groups including van der Waals interactions (Gattinoni and Michaelides, 2015). In order to discuss the solvent effect, molecular dynamics (MD) (Zeng, 2011), a combination of DFT with molecular dynamics (Wang et al., 2012, Guo et al., 2017), and combination of MO simulation with molecular dynamics are used (Khaled 2008, Khaled 2009). In all these simulations, adsorption of a single or small number of additive molecules are discussed.

Although the MD simulation is useful to discuss the dynamic process, such as diffusion coefficients in the many molecules systems (Hassan, 2016), this method was a subsidiary tool, since the charge transfer is unable to treat in usual MD. Recently, a force field called ReaxFF (van Duin et al., 2001), which can treat a chemical reaction with van der Waals and a long-range Coulomb interaction, is growing by including parameters for many elements (Senftle et al., 2016).

In this paper, we use the ReaxFF MD simulation to analyze the mechanism of the adsorption dynamics of the ensemble of the BTA anti-copper-corrosion additive molecules on the surface of copper. Most of the previous researches are focused on adsorption of single BTA molecule on the surface, or single BTA molecule with solvents, we can discuss the adsorption process of the molecular ensemble of the BTAs. By using ReaxFF, we can analyze not only the physical adsorption but the chemical adsorption process. The objective of this paper is to clarify the mechanism of the selective adsorption on the newly formed copper surface by analyzing the dynamics of the ensemble of the BTA molecules to the copper surfaces.

Our simulation shows two new aspects of the surface adsorption. The first one is the adsorption, surface diffusion, and aggregation of the BTAs on the copper (II) oxide (CuO) slab substrate surface. In this simulation, the physical adsorption and aggregation of many molecules are discussed. The second is the selective adsorption process of the BTAs on the copper (I) oxide ($Cu_2O$) and pure copper (Cu) hybrid surfaces. In this simulation, both the physical adsorption and the chemical adsorption, charge transfer between the slab and the BTA molecules, and aggregation of many molecules are discussed. This simulation gives insight into the long standing question of why the anti-copper-corrosion additive protects the newly formed surface.

## 2.0    SIMULATION PROCEDURE

We simulate and visualize the adsorption process of BTA molecules onto the copper surface by a MD simulation and analyze the dynamic behavior of the adsorbed molecules. Molecular dynamics is a simulation technique to provide the information about the micro structure and chemical and physical properties. The information comes from solving the time integral of the equation of motion for many particle systems and obtain trajectories of the particles. For input information, the potential function of interacting atoms, temperature, and pressure are used. Under a given condition, the atoms move toward the distribution in thermal equilibrium from solving the equation of motion of many atom systems.

As a potential function of interacting atoms, we use the Reax Force Field (ReaxFF) potential (van Duin et al., 2001). The ReaxFF potential is able to represent the formation and cleavage of bonds, because the parameter comes from quantum chemical calculations. In addition, the ReaxFF potential is compatible with a lot of elements (Senftle et al., 2016). The parameters for Cu and organic molecules are taken from the paper of Hu et al. (Hu et al, 2015) and their supplementary information. To calculate the time integral of the copper (copper oxide) and anti-copper-corrosion additive, we use molecular dynamics calculations software LAMMPS (Large-scale Atomic/Molecular Massively Parallel Simulator, Plimpton, 1995). LAMMPS is an open source that is good at efficient execution of the time integral execution and highly diverse.

In this paper, we simulate two models. The first model includes a copper (II) oxide (CuO) slab and BTA. In this model, we analyze the mechanism of the physical adsorption of BTA molecules on the surface and aggregation due to the surface diffusion. Since the copper atoms are almost polarized in this system, we believe aggregation without chemical adsorption can be discussed. The second model includes pure (not oxidized) copper (Cu) and copper (I) oxide ($Cu_2O$) in one plate, and BTA. Using this hybrid surface model, we analyze the mechanism of the selective chemical adsorption of the BTA molecules. Since the copper (I) oxide is the dominant structure in real phenomena, and the copper atoms can provide charges to the adsorbed additive molecules, thus copper (I) oxide is chosen. We describe the details of both models used in these simulations as follows.

## 2.1    METHOD FOR THE FIRST SIMULATION (CuO AND BTA)

To represent the dynamics of the adsorption process of the anti-copper-corrosion additive molecules onto the surface of copper (II) oxide (CuO), we first model a single BTA molecule as an anti-copper-corrosion additive and copper oxide slab. The bottom atoms of the CuO slab are fixed. Each part is stabilized using the ReaxFF potential (van Duin et al., 2010). The charge equilibration method (Rappe & Goddard III, 1991) is used for calculating the partial charge. We then preform molecular dynamics calculations with modeled anti-copper-corrosion additive molecules and CuO slab in the same space. The size of the simulation cell is (x, y, z) = (4.6837, 3.0347, 12.2) nm and a periodic boundary condition is used for the x and y directions. The number of Cu and O atoms is both 960 atoms. For the anti-copper-corrosion additive molecules, we disposed 20 BTA molecules in a random order. The setup condition of the temperature is 300 K using the Nose-Hoover thermostat, the time step size is 0.1 fs, and the number of time steps is 2,000,000. Based in these setups, the snapshot of the initial state of the model is shown in Figure 1.

In this simulation, we discuss the simulation results of the point BTA molecules absorbed from the aspect of their orientation to analyze the mechanism of the adsorption and the properties of the anti-copper-corrosion additive.

In order to analyze the orientation, we calculated the order parameter function f defined by equation (1).

$$f = \frac{3<\cos^2\theta> - 1}{2} \quad (1)$$

where $\theta$ is the angle between the focused axis and vertical vector from the surface of the BTA molecule and <> denotes the statistical average.

When the BTA molecule is non-oriented, the coefficient of orientation $<\cos^2\theta>$ will be 1/3 which is the statistical average value. If the surface of the copper slab and the surface of the BTA molecules are vertical (a), $\theta$ will be 0° or 180°, thus $<\cos^2\theta>$ equals 1. If the surface of the copper slab and the surface of the BTA molecule are parallel (b), $\theta$ will be 90° or 270°, thus $<\cos^2\theta>$ equals 0. Therefore, if the surface of the copper slab and the surface of the BTA molecule are vertically oriented, the coefficient of orientation f equals 1, and if the BTA molecule is randomly non-oriented, f is 0, and if the surface of the copper slab and the surface of the BTA molecule`s surface are parallel, it will be 0. Thereby, the lower the value, the more surface of the copper substrate and the surface of BTA molecule become vertical.

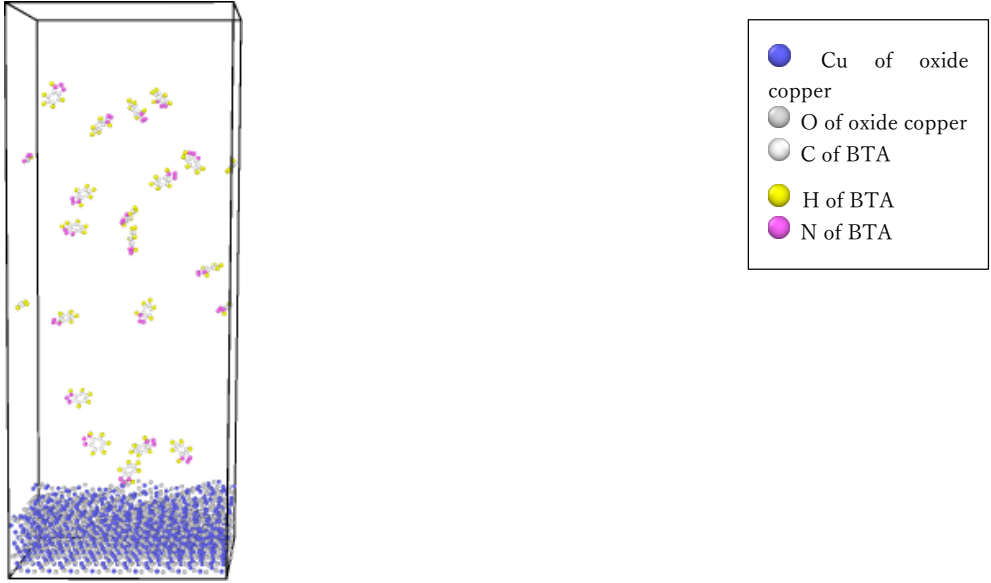

Figure 1: Snapshot of initial distribution of the first simulation.

## 2.2    METHOD FOR THE SECOND SIMULATION (Cu-Cu$_2$O AND BTA)

To prepare for the initial condition (Figure 2), we first modeled the copper (Cu), oxide copper (Cu$_2$O), and the anti-copper-corrosion additive. Each one was stabilized using the ReaxFF potential the same as in the first simulation. We then combined the copper (Cu) and oxide copper (Cu$_2$O) in the same simulation cell. The size of the simulation cell is (x, y, z) = (6.4924,  3.2535, 11.6) nm. The number of atoms for the slab models are set not to make a large dislocation. The number of Cu atoms in the Cu area and Cu$_2$O area are 1,296 and 784, respectively. The lowermost atoms are fixed. We finally preform the molecular dynamics calculation with modeled anti-copper-corrosion additive and slab in same simulation cell. For anti-copper-corrosion additives, we dispose 60 BTA molecules in a random order. The setup condition of temperature is 300 K, the time step size is 0.1fs, and the number of steps is 3,000,000.

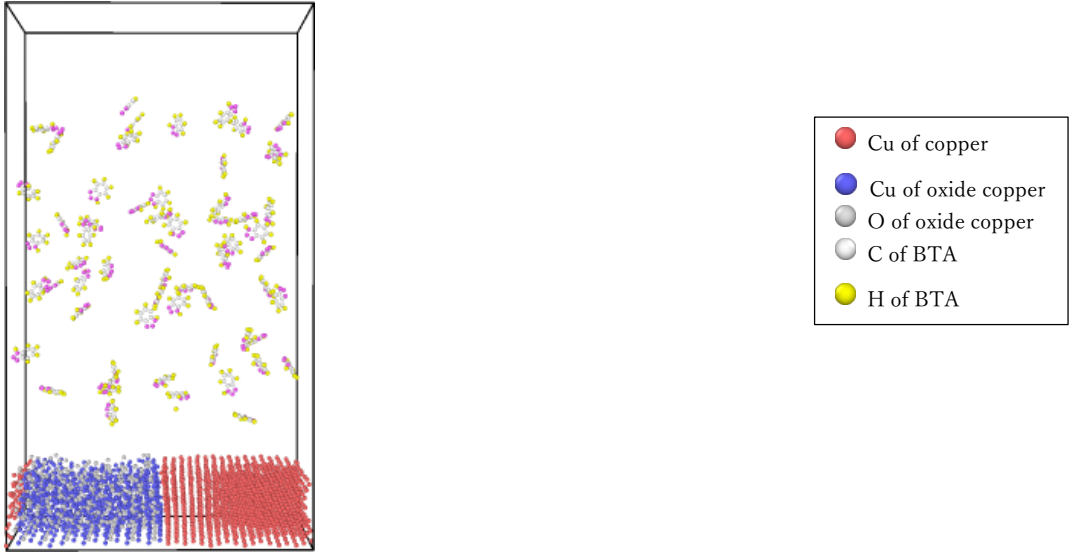
Figure 2: Snapshot of initial distribution of the second simulation.

## 3.0    RESULTS AND DISCUSSION

In both simulations, the BTA molecules are all adsorbed on the slab surfaces and show a quasi-equilibrium state, i.e., the total energy of the system converges to a single value and fluctuates due to thermal motion.

## 3.1    RESULTS OF THE FIRST SIMULATION (CuO AND BTA)

In the first simulation, from the initial state shown in Figure 1, we simulate the process of adsorption of the BTA molecules onto the surface of the copper (II) oxide (CuO). Figure 3 shows snapshots of the adsorption process of the BTA molecules onto the CuO surface. In 100 ps, 90 % of the BTA molecules are adsorbed and in 200 ps, all the BTA molecules are adsorbed. From the trajectory of the molecules in this process, the adsorption phenomenon is divided into 3 types. Figure 4 shows a schematic snapshot of each type. In the first type, the BTA molecule adsorbs onto the surface of CuO alone (Figure 4a). In the second type, the BTA molecules first form a group in a vacuum then adsorb onto the CuO surface (Figure 4b). In the third type, each BTA molecule first absorbs onto the surface, then forms a group by surface diffusion (Figure 4c). In addition to these types, after the absorption onto the surface of the CuO, the BTA molecules do not stay on the same place, but move on the surface (Figure 5).

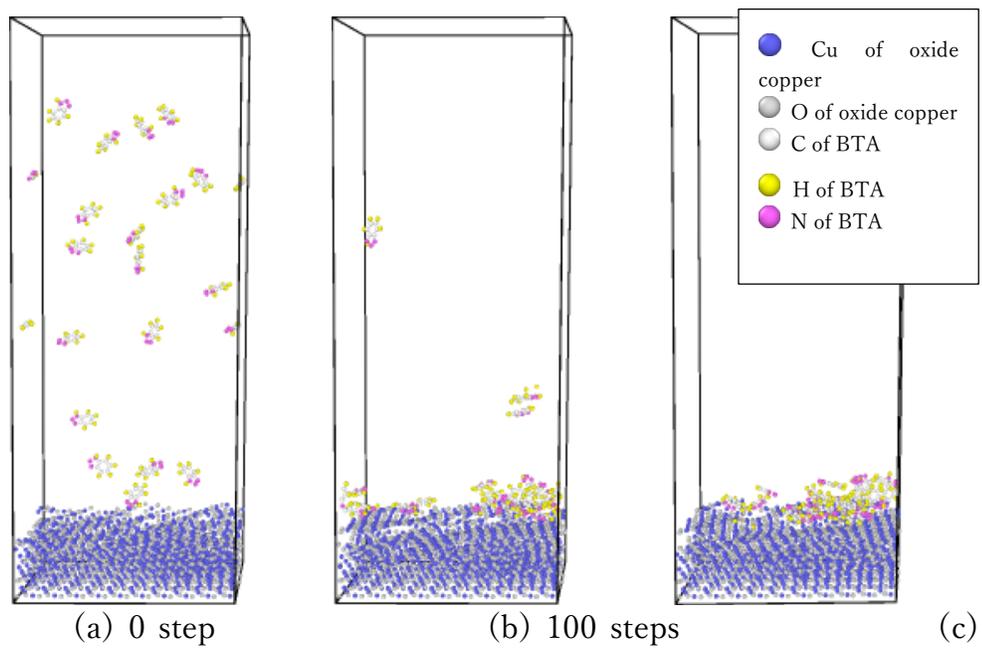

Figure 3: Snapshots of the adsorption process of BTA molecules onto the surface of CuO in the first simulation.

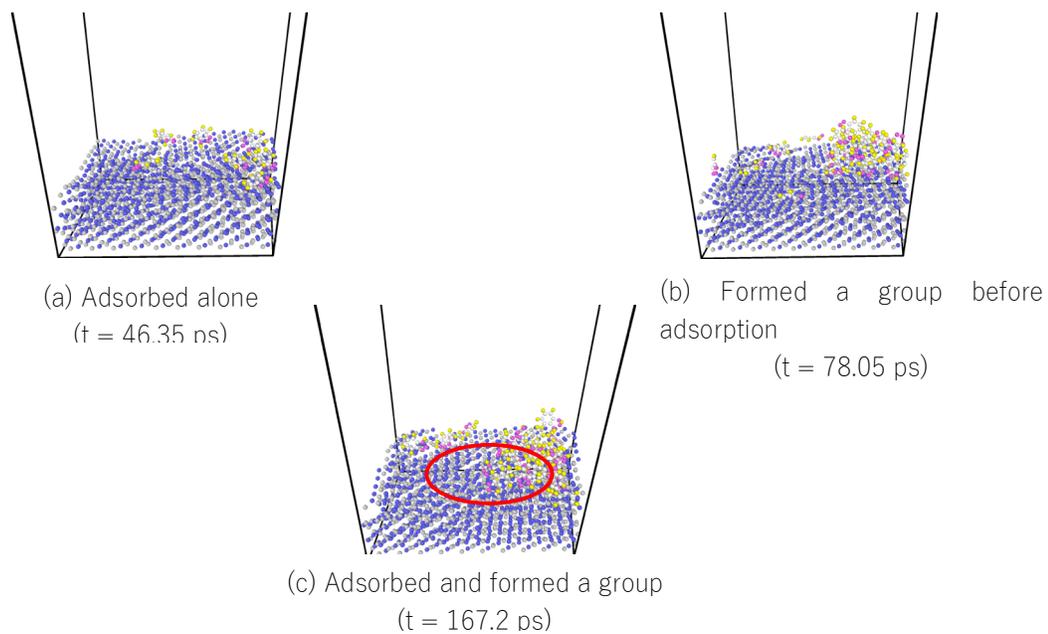

Figure 4: schematic snapshot of each adsorption type.

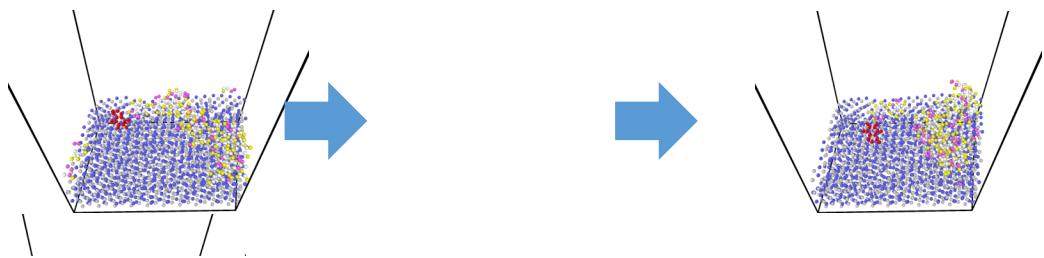

(t = 123.85, 156.10, 166.65 ps, from left to right).
Figure 5: Snapshots of a molecule moving on the surface by surface diffusion.

We analyzed the trajectory of the anti-copper-corrosion additive BTA molecules and determined the mechanism of adsorption. From the snapshot of the last step shown in Figure 3c, the BTA molecules that adhere to the copper surface make a layer a parallel to the surface. In order to analyze this process, we compare molecules that absorb onto the copper surface alone (type 1) and molecules that form a group before adsorbing (type 2). The average of the orientation factor at each time step is plotted in Figure 6 for the two types. The number of molecules is 3, for the type that absorbs alone (type 1, Figure 6a), and 11 molecules for the type which form a group before adsorption (type 2, Figure 6b).

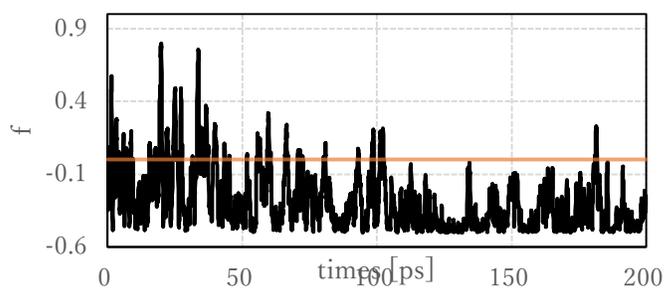

(a) Average of 3 molecules for type 1: absorbed alone.

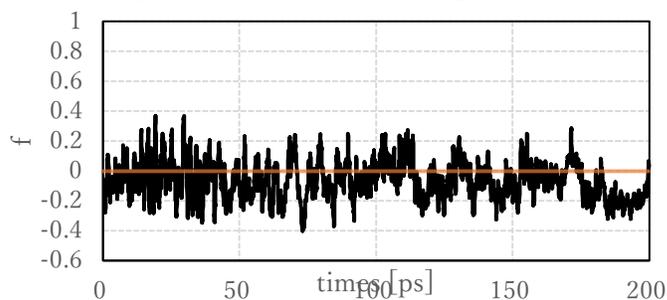

(b) Average of 11 molecules for type 2: form a group before adsorption.

Figure 6: Time development for average of the orientation factor in each time step. In both cases, the averages of the orientation factor are near 0 before the absorption (about 50 ps), which indicate that the orientation of the molecules is random. In type 1 shown in Fig. 6a, as the number of adsorbed molecules increase as the time step increases, because the parallel adsorption structure is dominant, the orientation factor shows a negative value. In type 2 shown in Figure 6b, the orientation factor fluctuates around 0, which means the direction of the BTA molecules is random after the adsorption.

We could not find the DFT calculation of the stable structure of the BTA molecule on the CuO surface, therefore, the result of the MD simulation cannot be compared to the other theoretical studies. However, we can understand the MD result as followings. In the CuO surface, the electrons of the copper atoms totally moved to the oxygen atoms, thus the nitrogen atoms in the BTA do not obtain electrons from the copper atoms. This means that the adsorption is a physical adsorption. If the inter-molecular interaction, such as the van der Waals and Coulomb, are dominant to the adsorption process, the face-to face structure may be stable, as have been shown in classical MD simulation of adsorption on Fe surface by Guo et al., 2017. The random orientation shown in the type 2 (Figure 6b) can be understood by the excluded volume effect between the BTA molecules. At room temperature, the BTA molecules show an amorphous-like structure on the surface. One of the newest experimental research (Zhao et al., 2018) shows that the CuO is the intermediate in the anti-corrosion process and Cu(I) BTA is dominant rather than Cu(II)BTA, which is consistent with our results that the BTA physically adsorb on the Cu(II) O surface.

## 3.2    RESULTS OF THE SECOND SIMULATION (Cu-Cu$_2$O AND BTA)

From the initial state shown in Figure 3, we simulate the adsorption process of the BTA molecules onto the surface of the copper (Cu) or copper (I) oxide (Cu$_2$O) slab. The snapshots of the time evolution of this simulation are shown in Figure 7.

In this simulation, we investigated the trend onto which area the BTA molecules adsorb. The number of BTA molecules adsorbed onto the copper (Cu) or copper (I) oxide (Cu$_2$O) is shown in Figure 8. At the beginning of the simulation (0 ps to 40 ps), the number of BTA molecules increased as the time increased on both sides. This is because the adsorption process is a physical adsorption and is governed by a random process. The number of BTA molecules on the Cu area then quickly increased and the increased ratio decrease in the Cu$_2$O area. Note that the decrease in the number of adsorbed BTA molecules on the Cu$_2$O area is around 40 ps. This is due to both the removal from the surface of the Cu$_2$O area to the vacuum, and movement from the Cu$_2$O area to the Cu area by surface diffusion. After this turning point, the number of BTA molecules on the Cu area quickly increased due to the long-range molecular interactions between the adsorbed molecules and the molecules in the vacuum. The surface diffusion from the Cu$_2$O area to the Cu area also increased the number of molecules in the Cu area. Finally, about a 5 times greater number of BTA molecules was adsorbed on the Cu area.

In addition, the binding part of each BTA molecule to the surface of the molecules in the Cu area seemed different from that in the Cu$_2$O area. As shown in Figure 7b, during the initial process, the imino-group (N atom) of BTA molecules comes lower side when it adsorbs onto the Cu area than the Cu$_2$O area. Both experiments and the theoretical analysis which we have shown in the introduction of this paper, have proven that in many cases, the imino-group face to the copper surface and the charge transfer occur. The percentage of the molecules which imino-group is bound on the slab atoms in both areas is shown in Figure 9. During the process of the initial adsorption (0 to 150 ps), the binding part of the BTA molecules are different in the Cu area and the Cu$_2$O area. In case of BTA molecules adsorb onto Cu$_2$O, the binding part of initial adsorption is not specific. Whereas in case of BTA molecules adsorbed onto Cu, the percentage of BTA molecules which N atoms are bound is high. This is because the initial adsorption process between Cu surface and BTA molecules is thought to be due to the transfer of charge between the Cu atom and N atom in the BTA. After the initial adsorption process, in both areas, the majority of the binding part is N atoms in the BTA. This is because the Cu atoms in both areas are able to give electrons to N atoms, and the molecules are chemically adsorbed.

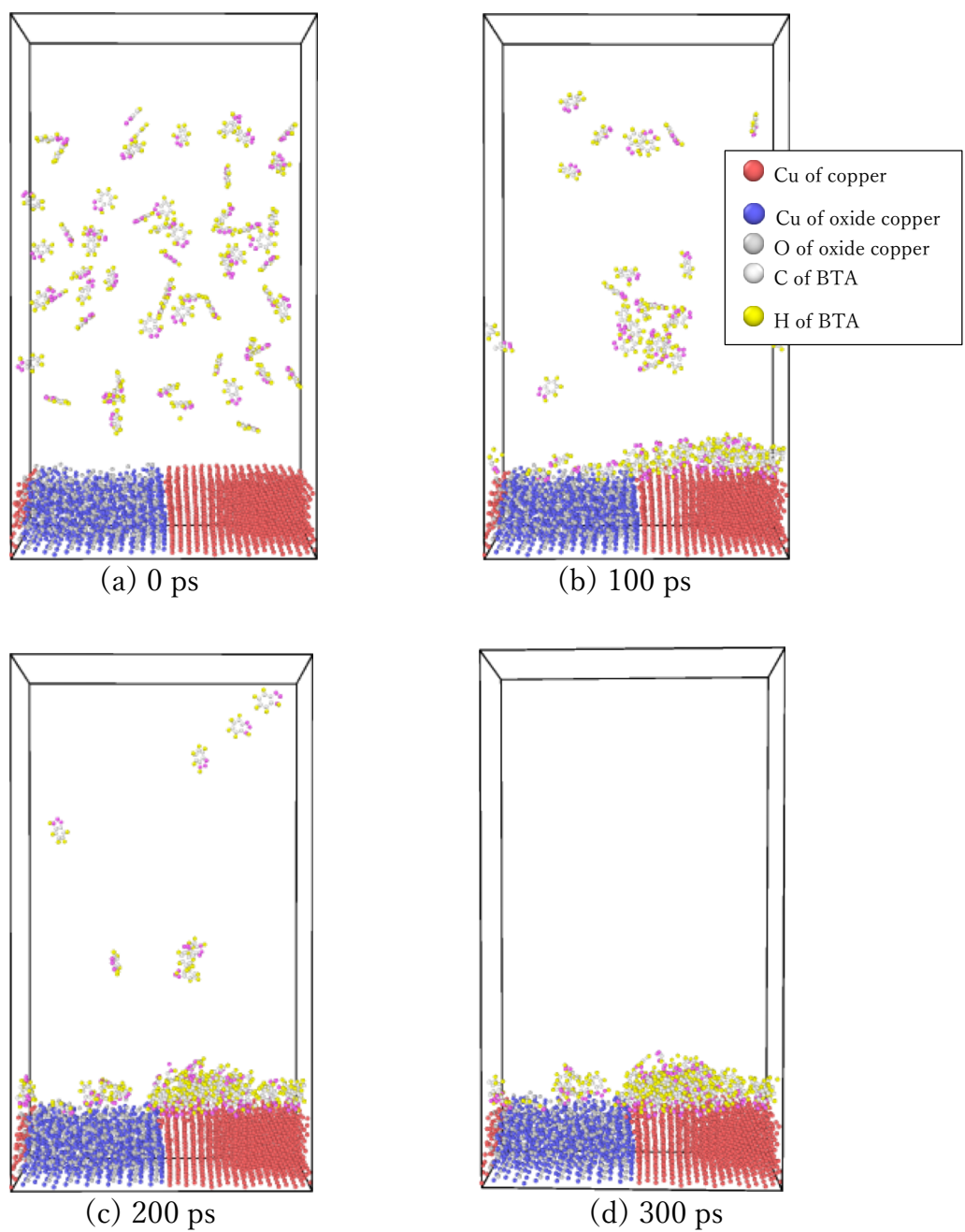

Figure 7: Snapshots of the adsorption process of BTA molecules onto the surface of Cu$_2$O and Cu hybrid surface in the second simulation.

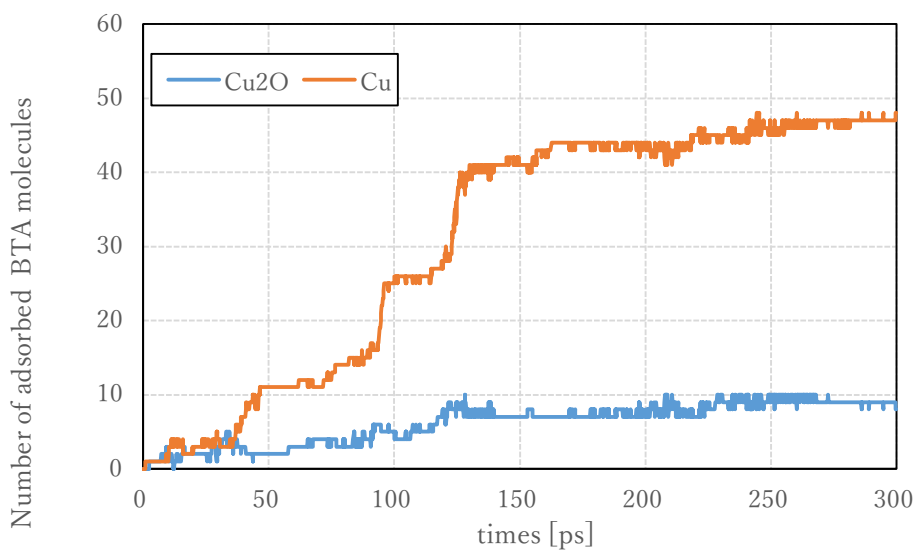

Figure 8: Number of BTA molecules adsorbed onto the copper (Cu) area or the copper (I) oxide ($Cu_2O$) area.

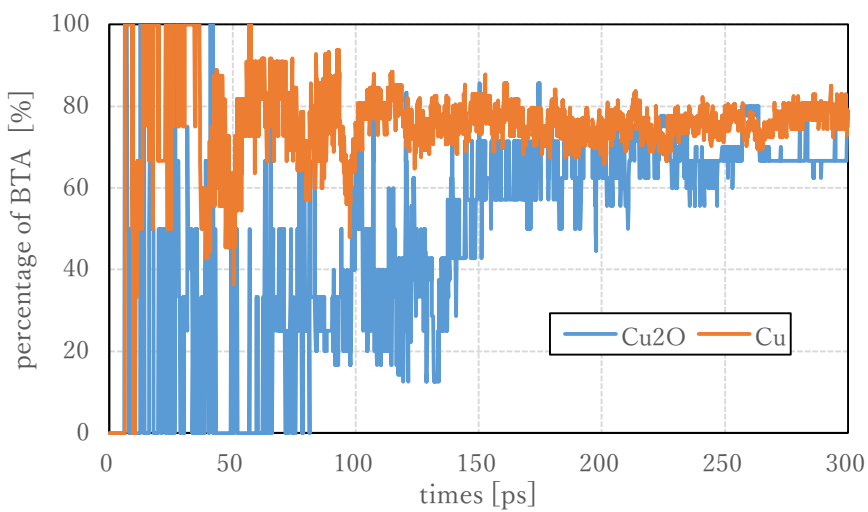

Figure 9: Percentage of BTA molecules in which the imino-group is bound to the slab atoms in the copper (Cu) area or the copper (I) oxide (Cu2O) area.

In order to understand the dynamics of the charge transfer, we plotted the time development of the averaged charge of the copper atoms in the top layer of the slab and the middle of the slab in each area (Figure 10). Initially, the copper atoms in the $Cu_2O$ area are already polarized, while the copper atoms in the Cu area are not.

After the simulation, the charge of the Cu atoms in the surface of the Cu area increases due to the charge transfer to the imino-group of the BTA molecules and become positive value. The charge in the $Cu_2O$ area slightly increases. This means the chemical adsorption occur in the $Cu_2O$ area too. However, the number of BTA molecules in the $Cu_2O$ area are restricted since the charge on copper atoms are already polarized, and there is a slight tendency to transfer the charge to the adsorbed molecules.

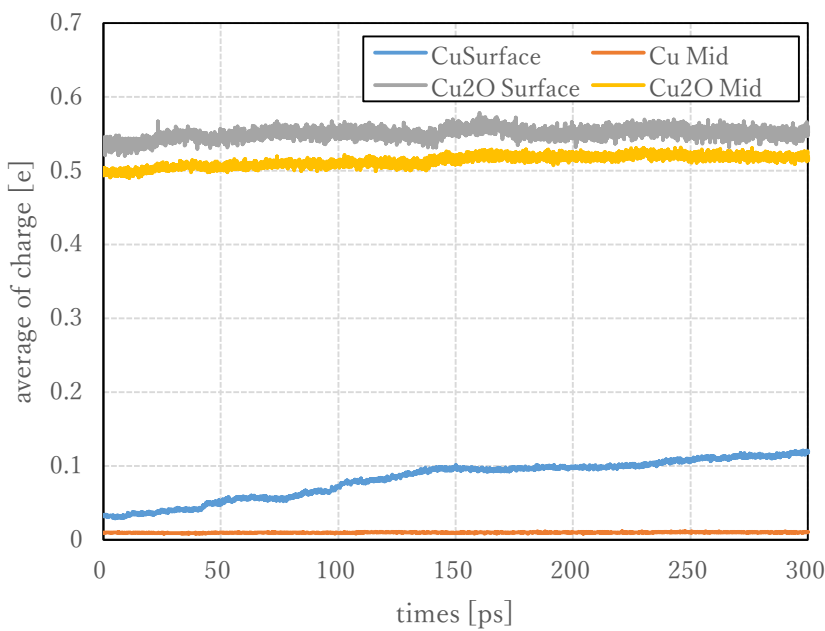

Figure 10: Time development of averaged charge of the copper atoms in the top layer of the slab and the middle of the slab in the copper (Cu) area or the copper (I) oxide ($Cu_2O$) area.

Comparing the final snapshot of the two simulations (Figure 3c and Figure 7d), one can distinguish that the BTA molecules are standing in both areas in Figure 7d and they are laying in Figure 3c. To analyze this difference, we calculated the order parameter function of the BTA molecules which adsorb onto the Cu area.
Figure 11 shows the orientation of the first layer of the BTA molecules adsorb onto copper plate. In the first simulation, we calculated the orientation of the BTA molecules adsorbed onto the copper (II) oxide (CuO), and the result of orientation was parallel with the surface (Figure 6b). Although the both f values are almost 0, the average in Figure 11 is higher than that shown in Figure 6b. This is due to the oriented adsorption, which is consistent with a previous theoretical prediction using DFT (Kokalj et al., 2010). In our simulation, the reason why the orientation factor decreases from 1, is the number of adsorbed molecules is high and making a complex structure under the realistic thermal effect.

We next calculated the amount of charge transfer of each of the atoms belonging to the BTA molecules to figure out the precise mechanism of the BTA molecule adsorption onto the copper slabs.

We first calculated an average of the charge transfer amount of the N atoms and H1 atom (Figure 12) belong to imino-group of the BTA molecules which are the absorbed point BTA molecules. These results are shown in Figure 13. An average of the amount of charge of the N atoms (Figure 13a) does not show a large difference between the BTA molecules adsorbed onto the Cu area and the $Cu_2O$ area, however, the average of the Cu area is slightly greater than that of the $Cu_2O$ area. Comparing the charge of H1 atom (Figure 13b), the charge increases about 0.05 [e] in the $Cu_2O$ area, whereas it slightly increases in Cu area. In order to understand this result, we calculated the average of the charge of oxygen atoms exist on top layer and intermediate layer of the $Cu_2O$ area which BTA molecules adsorb and the charge of the H1 atom has changed (Figure 14). The charge on the surface of the O atoms decreases as the adsorption occur. This means that the charge transfer from the H1 atom to the O atoms occurs.

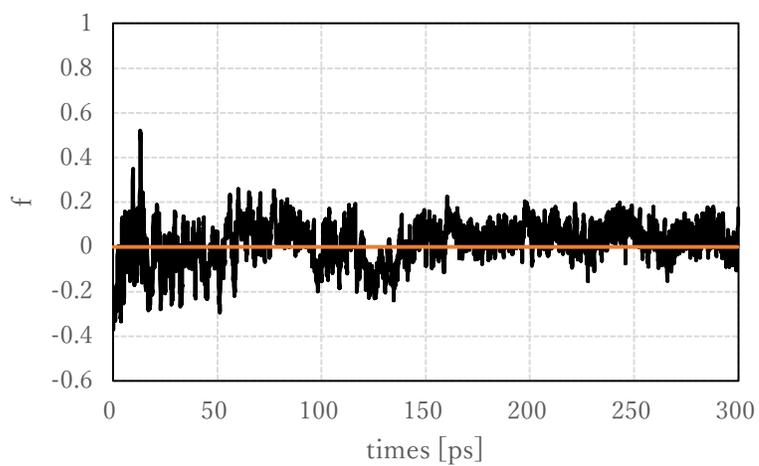

Figure 11: Time development of orientation function calculated from BTA molecules which adsorbed onto the copper area.

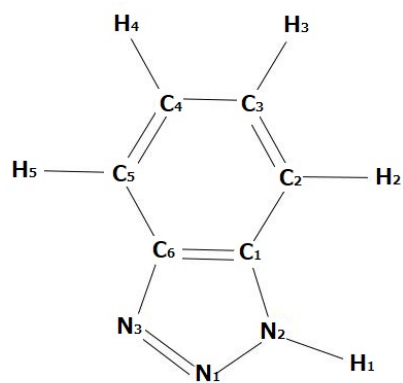

Figure 12: Chemical structure and the number of atoms in this simulation of the BTA.

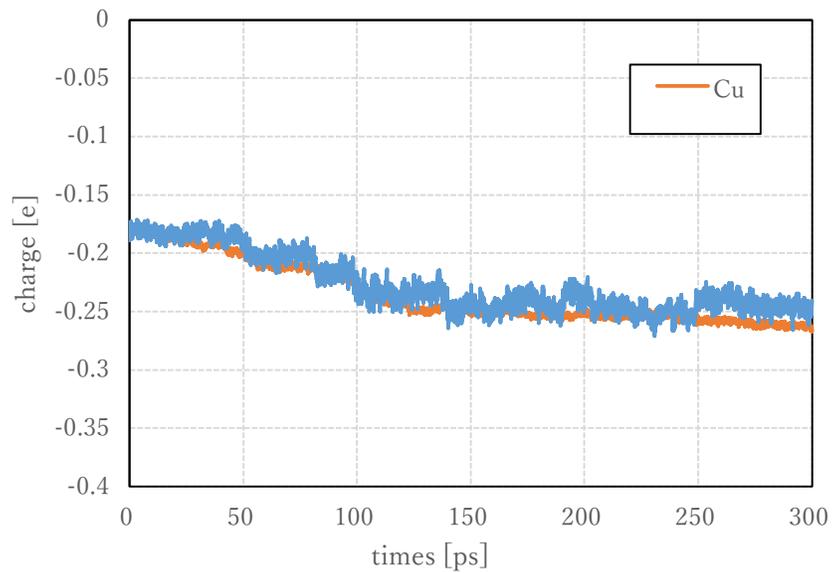

(a) Charge of N atoms.

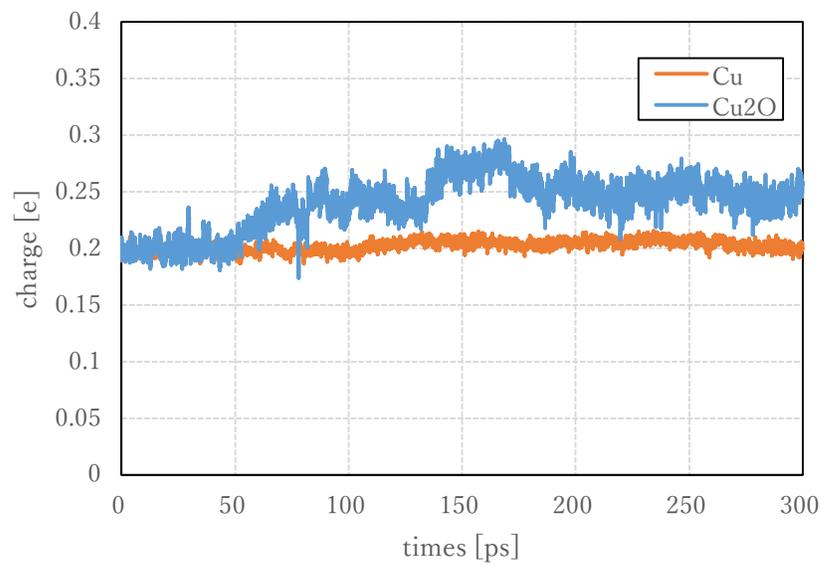

(b) Charge of H1 atoms.

Figure 13: Time development of the average of the charge on each BTA molecule during the simulation.

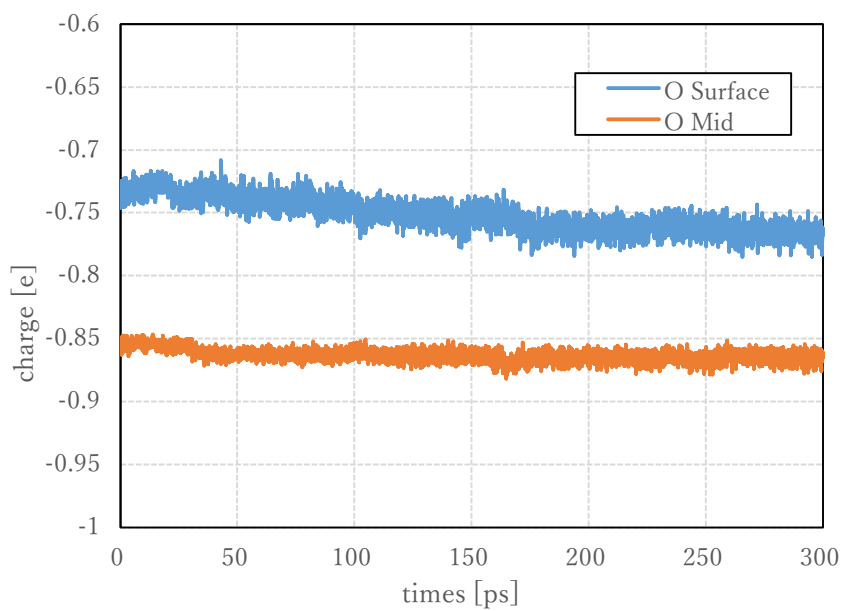

Figure 14: Time development of the average of the charge of O atoms present on the surface and middle of the copper (I) oxide ($Cu_2O$) area.

We finally summarize the details of each charge of the atoms belonging on the BTA molecule, before and after this simulation (Figure 15). The blue atom indicates that the atom is charged negatively, and red indicates a positive charge. Also, the deeper color means a higher value. The result is consistent with the partial charge distribution of previous DFT calculations (Jiang et al., 2004) or MO calculations (Zeng, 2011), at least qualitatively. A slight increase in the positive charges of the hydrogen atoms in the benzene ring during the adsorption process may due to the charge normalization effect in each molecule.

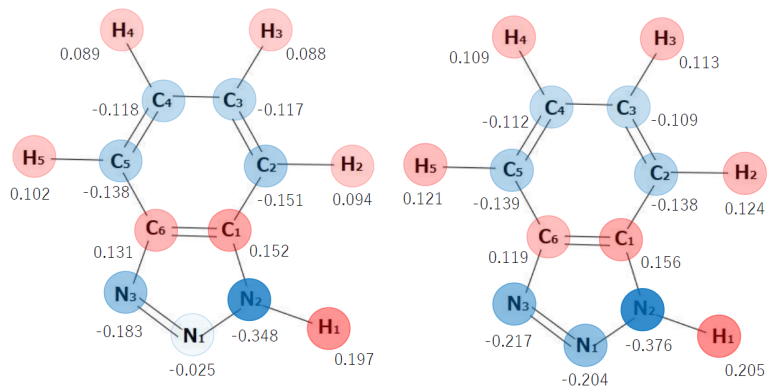

(a) BTA molecule which adsorb onto the copper (Cu) area.

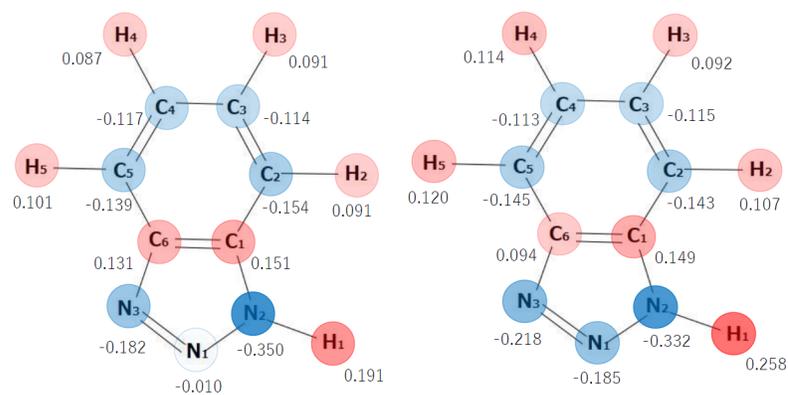

(b) BTA molecule which adsorb onto the copper (I) oxide ($Cu_2O$) area.

Figure 15: Average of charge of each atom belonging to the BTA molecule. At the initial state (t = 0, left) and at the final state (t = 300 ps, right).

Based on these analyses, let us explain the difference in the adsorption amount between the Cu area and the $Cu_2O$ area shown in Figure 8. The copper atoms and N atoms of the BTA molecules are exchanging their charge, so this type of adsorption between the copper plate and BTA molecules is thought of as chemical adsorption in both the Cu and $Cu_2O$ areas. Since the N atoms of BTA molecules that adsorb onto the $Cu_2O$ area are already negatively charged, and the O atoms of the $Cu_2O$ area are also negatively charged, the BTA molecules in the $Cu_2O$ area are difficult to physically adsorb due to the repulsive interaction between these atoms. Only when the N atoms are on the Cu site, and the H1 atoms are on the O site is the adsorbed structure stable, which is consistent with the previous DFT calculation (Jiang et al., 2004).

Compared with the $Cu_2O$ area, the BTA molecules are easily adsorbed onto the Cu area, because the Cu atoms, which exist on the top layer of the $Cu_2O$ area, are originally polarized, and that on the top layer of the Cu area are not yet charged. In spite of this difficulty, if the conditions to adsorb are matched, chemical adsorption will occur in the $Cu_2O$ area by changing charges between the Cu and N atoms of the BTA molecules.

Due to the surface diffusion of the physically adsorbed BTA molecules on the $Cu_2O$ area, the BTA molecules chemically adsorb when the molecules reach the Cu area. The chemical adsorbed molecules changes its permanent dipole moment due to the charge transfer from the Cu atoms in the slab. This effect enhances the adsorption between the molecules in the vacuum area and the molecules already adsorbed. The more polarized BTA molecules attract the other molecules to the surface. Although the chemical adsorbed molecule in the $Cu_2O$ area also attracts the other molecules in a vacuum, the total effect due to the already adsorbed molecules is different in both areas. These two effects, i.e., surface diffusion and polarization effect, make the turning point shown in Figure 8, more critical. Finally, the number of BTA molecules becomes very different.

Every simulation study should be related to the experimental results. Our simulation seems that we succeeded in explaining the most important question why the anti-copper-corrosion additive protects the newly formed surface. Even if the mass of the anti-copper-corrosion additive is low in an aqueous or oil solution, the additive molecules show a selective adsorption process to find the newly formed surface, and not the oxidized surface. This phenomenon is found by calculating the many number of molecules with a charge transfer between atoms. Although a more precise method, such as the ab initio MD, will provide more detailed results, we think the results in this paper will not qualitatively change.

For the further work, many aspects must be studied. Since the real phenomena occur in the liquid phase, the effect of the solvent must be studied. The solvent is treated by classical molecular dynamics, and as in many previous studies, the results will change in the electrolyte solution or in a high or low pH aqueous solution. The results in this paper will slightly change in the oil solution, however, we need to change the additive molecule which has a long hydrocarbon chain to be soluble. In such further simulation, the ReaxFF may still be useful by the feature explained above.

## 4.0  CONCLUSIONS

Two models of the adsorption process of the anti-copper-corrosion additive, benzotriazole (BTA), onto the surfaces of a copper (II) oxide (CuO) slab, and hybrid slab with copper (Cu) and copper (I) oxide ($Cu_2O$) are simulated using molecular dynamics with the ReaxFF potential which includes the charge calculation.

From the result of the first simulation, adsorption between the BTA molecules and copper oxide surface, the BTA molecules form an adsorbed layer in parallel on the surface of the CuO. The BTA molecules which adsorb alone, show a surface diffusion due to the weak physical adsorption. In addition, some of BTA molecules aggregate on the surface.

From the result of the second simulation, a selective adsorption phenomenon is found. The number of BTA molecules adsorbed on the Cu area is 5 times greater than that on the $Cu_2O$ area. An analysis of the time evolution of the atomic charge showed that in both areas, the BTA molecules finally undergo a chemical adsorption. The partially charged nature of the $Cu_2O$ and the originally charged Cu atoms make the adsorption less than in the Cu area, and as a result, the major part of the BTA molecules aggregate on the Cu area.

In real phenomena, the reason why a few anti-corrosion additives are able to protect a metal surface is thought to be the selective nature of the adsorption onto the newly formed metal surface.

ACKNOWLEDGEMENT
The authors gratefully acknowledge Dr. Kunichika Kubota, Dr. Junpei Yoshida, Dr. Takafumi Ishii, Dr. Takao Kudo, and Prof. Masabumi Masuko for their useful discussions. This study was supported by JSPS (Japan Society for the Promotion of Science) KAKENHI (Grants-in-Aid for Scientific Research) Challenging Research (Exploratory) Grant Number 18K18813.